\newcommand{\Hlogic}{\hat{H}}
\newcommand{\Hlogicf}{\Hlogic_p}
\newcommand{\Hlogici}{\Hlogic_i}
\newcommand{\Alogic}{A}
\newcommand{\Blogic}{B}
\newcommand{\paulilogic}{\hat{\sigma}}

\newcommand{\Hphys}{\tilde{H}}
\newcommand{\Hphysf}{\Hphys_p}
\newcommand{\Hphysi}{\Hphys_i}
\newcommand{\Hphyscplaq}{\Hphys_c}
\newcommand{\Aphys}{\tilde{A}}
\newcommand{\Bphys}{\tilde{B}}

\newcommand{\pauliphys}{\tilde{\sigma}}

\documentclass[aps,pra,twocolumn,final,floatfix,superscriptaddress,longbibliography,10pt]{revtex4-2}

\usepackage[utf8]{inputenc}
\usepackage{physics}
\usepackage{mathtools}
\usepackage{dsfont}
\usepackage{amsmath}
\usepackage{xcolor}
\usepackage{tikz}
\usepackage{soul}
\usepackage{bbold}
\usepackage{graphicx}
\usepackage[export]{adjustbox}
\usepackage[breaklinks=true,colorlinks,citecolor=blue,linkcolor=blue,urlcolor=blue]{hyperref}

\usetikzlibrary{graphs,graphs.standard}

\begin{document}

\title{A Set of Annealing Protocols for Optimized System Dynamics and Classification of Fully Connected Spin Glass Problems}

\author{Gino Bishop}
\affiliation{Theoretical Physics, Saarland University, 66123 Saarbr{\"u}cken, Germany}
\affiliation{Institute for Quantum Computing Analytics (PGI-12), Forschungszentrum J\"ulich, 52425 J\"ulich, Germany}
\affiliation{Mercedes-Benz AG, Stuttgart, Germany}  

\author{Simone Montangero}
\affiliation{Dipartimento di Fisica e Astronomia, Universit\'a degli Studi di Padova, 35131 Padova, Italy}
\author{Frank K. Wilhelm} \affiliation{Theoretical Physics, Saarland University, 66123 Saarbr{\"u}cken, Germany}\affiliation{Institute for Quantum Computing Analytics (PGI-12), Forschungszentrum J\"ulich, 52425 J\"ulich, Germany}

\date{\today}

\begin{abstract}

We perform exact diagonalization and time evolution of the Lechner-Hauke-Zoller (LHZ) annealing architecture [Science Advances 1(9), e1500838 (2015)] for ten \textit{physical} qubits.
Thereby, on a training set consisting of $2400$ problem instances, we perform the optimization task of tuning the local fields with the goal to identify a set of fixed optimal annealing protocols, that outperforms linear protocols on a large class of arbitrary LHZ problem instances. We show that average ground state fidelities of $0.9$ can be achieved by applying optimized protocols onto groups of problem instances with similar energy landscapes. Particularly, the set of optimized annealing protocols reduces annealing time required to reach a predefined threshold ground state fidelity by an average of $\sim72\%$, corresponding to a speed-up of factor $\sim3.5$. Moreover, as these protocols are meant to be readily applicable in experimental setups, they can be used as a starting point for problem-specific protocol optimization. In reverse, we discuss how previously optimized protocols can potentially be used to gauge the instantaneous energy landscape of a spin glass problem. Albeit simulations were performed on the LHZ architecture, identification of optimized protocols is not limited to either simulations or local connectivity only.

\end{abstract}

\maketitle

\section{Introduction}
\label{sec:intro}
Quantum annealing is a technique used for finding the global minimum of a given objective function. The nature of these objective functions are often combinatorial optimization problems --- prominent examples are the traveling salesman problem \cite{inbook} and the maximum cut problem \cite{maxcut}. In the context of quantum computing, combinatorial problems can be cast into the form of a transverse-field Ising model, where the goal is to find the minimum energy configuration of spins. With growing system sizes it becomes more and more difficult to solve such systems classically~\cite{Albash2018} as memory requirements double with each additional spin. Hence the demand for sophisticated quantum algorithm.

The transverse-field Ising model is described by a problem Hamiltonian $\hat{H}_p$, for which we seek to find the ground state $\lvert \Phi_p \rangle$. Typically, one performs a population transfer between two quantum states $\lvert \Phi_i \rangle, \lvert \Phi_p \lvert$, where $\lvert \Phi_i \rangle$ is the lowest energy eigenstate of $\hat{H}_i$, via the annealing protocols $A(\tau), B(\tau)$ as $\Hlogic{(\tau)} = \Alogic{(\tau)} \Hlogici + \Blogic{(\tau)}\Hlogicf$, with $\tau = t/T \in [0,1]$. In adiabatic quantum annealing \cite{born} the population of the instantaneous ground state at any time $\tau$ stays constant, given the minimum instantaneous energy gap $\min_\tau \Delta E(\tau) = \min_\tau \lvert \epsilon_m(\tau) - \epsilon_n(\tau) \lvert \equiv \Delta E$ between the two lowest instantaneous eigenstates $\lvert m(\tau) \rangle, \lvert n(\tau) \rangle$ is non-zero at all times. The drawback is its slowness, as the adiabatic condition \cite{Albash2018}
\begin{equation}
T \gg \max_{\tau\in [0,1]} \frac{\lvert \langle m(\tau) \lvert \partial_{\tau}\hat{H}(\tau) \lvert n(\tau) \rangle \lvert}{\Delta E^2}
\label{eq:adiabatic_condiction}
\end{equation}
states that experimental time $T$ scales at best as $\mathcal{O}(\Delta E^{-2})$ and at worst as $\mathcal{O}(\Delta E^{-3})$. Due to the slowness in adiabaticity, the protocols $A(\tau), B(\tau)$ are often subject to adaptations, providing a faster route to achieve results similar to adiabatic annealing \cite{Torrontegui2013}. Shortcuts to adiabaticity \cite{chenxi} include counterdiabatic driving protocols \cite{berry}, that prevent the system from leaping onto higher energy states. Alternatively, optimal control theory \cite{herschel} can be used to allow (and enforce) leaps to higher energy levels, whereby large overlap between simulated or experimental quantum state with the analytic ground state can be achieved even in a non-adiabatic regime. Consequently, recent studies include the automatization process of designing annealing protocols \cite{Chen2022} as well as optimizing them based on ground state fidelities \cite{herr2017optimizing}, \cite{Matsuura2021}. Lastly, not to forget the recent experiment on a D-Wave annealing device \cite{dwave} giving proof of the scaling behaviour of kinks within the Kibble-Zurek \cite{kibbler} framework for both open and closed quantum systems. 

While specialized annealers exist for not-fully connected graphs, the intriguing advantage of the Lechner-Hauke-Zoller (LHZ) ~\cite{Lechner2015annealer} architecture is the achievement of all-to-all connectivity via fully programmable local interactions. A one-to-one mapping between local and full connectivity is established via mapping $N$ \textit{logical} qubits onto $K=N\cdot(N-1)/2$ \textit{physical} qubits. Locally tunable constraints need to be introduced to balance out the increase in the amount of degrees of freedom. An implementation with Rydberg atoms in an optical lattice has been proposed ~\cite{Glaetzle2017rydberg}, where the key challenge is reportedly the implementation of the constraints. Lastly, a generalization of the LHZ scheme within the stabilizer mechanism is possible~\cite{Rocchetto2016stabilizers}. The authors show that their proposed stabilizer formulation can reduce the qubit count in specific problems as well as map arbitrarily high-order interactions in \textit{logical} spins onto single \textit{physical} qubits.

In this paper, we explore the original LHZ proposal with $N=5$ fully connected \textit{logical} qubits. In particular, we aim to contribute to the process of finding optimal annealing schedules by providing a set of fixed optimized protocols $S \equiv \{(A(\tau), B(\tau))_1, (A(\tau), B(\tau))_2, \dots\}$ applicable to arbitrary transverse-field Ising instances within the LHZ scheme. This is done by analyzing in depth a sample of $2400$ instances, which will serve as our training set. As these desired protocols are applied and evaluated on the basis of the large sample, by design protocols in $S$ are not optimal for any individual instance, but rather outperform linear ones on a large class of programmable problems. Moreover, the protocols are deliberately different from each other, so that the variety of all protocols cover a wide range of instances. Lastly, all protocols can further be used as a collection of guess pulses for thorough optimization of system dynamics in any underlying problem of similar size. For our purpose, it is sufficient to represent the protocols $(A(\tau), B(\tau))_i$ by a single protocol $s_i(\tau)$, such that $S = \{s_1(\tau), s_2(\tau), \dots\}$.

Optimizing annealing protocols enables us to find solutions for combinatorial optimization problems faster ~\cite{Caneva2011}, especially if we allow for detachment from the adiabatic regime. Prior use of optimization techniques is required to find optimal protocols $s_i(\tau)$. For this task, we chose the \textit{dressed Chopped RAndom Basis algorithm} (dCRAB)~\cite{crab, Rach2015} which is a bandwith-limited optimal control technique capable of exploring phase space in an ergodic fashion. It is part of the \textit{Quantum Optimal Control Suite} (QuOCS), \cite{QuocsRossignolo2023}.

The adiabatic condition, Eq. \ref{eq:adiabatic_condiction}, requires the simulation time $T$ to be sufficiently large to enable conversion between two quantum states, with $\Delta E$ denoting the minimum instantaneous spectral gap between the ground state and the first excited state. What is an appropriate magnitude of $T$ to use for an arbitrary instance? The answer becomes increasingly difficult to anticipate with more instances to be solved with the same strategy, not to forget that extracting $\Delta E$ is itself generally a challenging problem \cite{Albash2018}.

One way to handle this is to cluster instances according to some shared traits. We call the set that consists of groups of instances of similar kind $G \equiv \{g_1, g_2, \dots\}$. These kinds could include the magnitude of the global minimum gap $\Delta E$, the number of local minimum gaps and their distribution throughout the annealing process. Intuitively, instances that indeed share certain traits are then put into a single group $g_i$. We argue, that the assignment of instances to such groups simplifies the effort of finding optimal annealing protocols. Following this notion, we present routes to group instances, find optimal protocols for the individual groups $g_i$ and probe their performance on a test set. The goal is to find a suitable protocol $s_i(\tau)$ for every group $g_i$, i. e. $\lvert S \lvert = \lvert G \lvert$. 

The work is structured as follows. Part \ref{sec:intro} marks the introduction to the field of research. In the second part \ref{sec:model}, we provide a brief summary on the LHZ architecture and how we employ protocols to arbitrary Ising problems of five \textit{logical}, fully connected spins. The third part \ref{sec:optimized_schedules} consists of breaking down the generation of the set of problem groups $g_i$ distinguished by the average minimum energy gap. In parallel, we show the design process of optimized protocols for each individual group. Part four \ref{sec:results} deals with the application of $s_i(\tau) \in S$ to a test set consisting of groups of LHZ instances in order to showcase the speed up in system dynamics. Lastly, we provide an outlook on how to push forward the concept of applicable pre-optimized protocols and what to expect in larger systems, consisting of more than five fully connected \textit{logical} qubits. Hereby, we shift from a simulation frame to a laboratory frame and discuss a method to generate optimized protocols via annealing experiments.

\section{LHZ architecture}
\label{sec:model}

\begin{figure*}
\centering
\includegraphics[scale = 0.25]{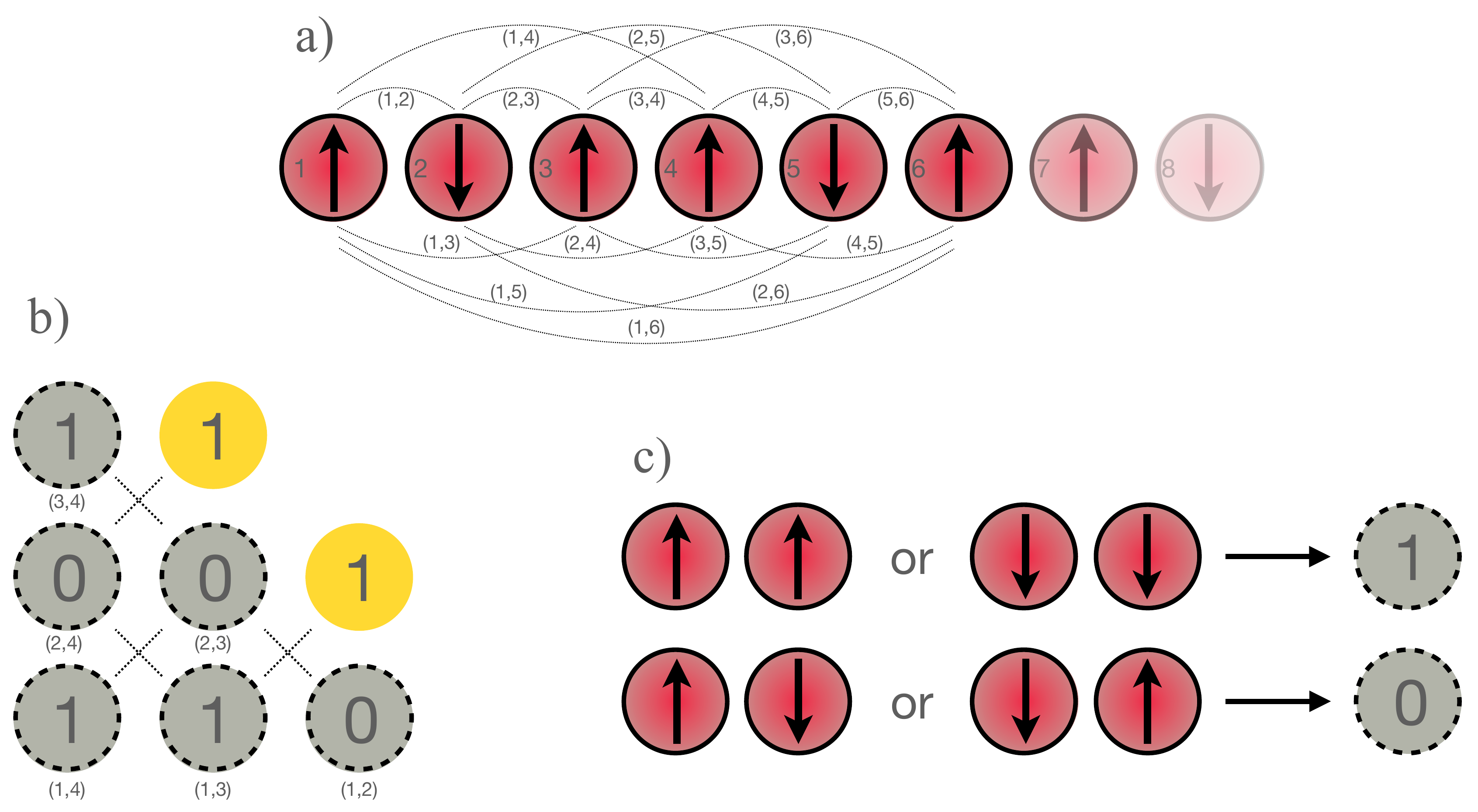}
\caption{(Color online) Mapping of the infinite-range interactions to local interactions as per ~\cite{Lechner2015annealer}. a) Ising chain with six \textit{logical} qubits (red, solid frame). Lines mark the interactions between qubit pairs $(i, j)$. b) Corresponding \textit{physical} qubit (gray, dashed frame) architecture with local interactions. Number of spin up in a given plaquette must be even. Two cornered qubits (yellow, open frame) are fixed to protect the constraints in edge cases. c) Set of rules for mapping \textit{logical} (red, solid frame) to \textit{physical} (gray, dashed frame) qubits.}
\label{fig:architecture}

\end{figure*}

We chose to work on the LHZ architecture for two main reasons. Firstly, albeit four-body constraints need to be implemented, which might pose a technical challenge, it is hardware friendly as interactions are implemented locally. Secondly, from~\cite{Lechner2015annealer} we learn that there are discrepancies between the Ising spin glass and the programmable model in terms of three key metrics: deviation of the lowest energy levels at $\tau=1$, the ratio of instantaneous gaps during the sweep and the time of occurrence of the minimum gap. Although these discrepancies can be damped depending on the choice of constraint strength $C$ against interaction strength $\lvert J \lvert$, they will not vanish utterly. It is thus straightforward to argue, that system dynamics will differ in both models. In this work, the authors aim to shed light onto the system dynamics within the programmable LHZ model. 

Following~\cite{Lechner2015annealer}, we give a brief overview on the LHZ architecture. For a thorough description we urge the interested reader to acknowledge the original, above mentioned paper.
\newline
We start by describing the Hamiltonians and the mapping from \textit{logical} to \textit{physical} spins.

\paragraph{Fully connected annealer:}

We start from a classical Ising spin-glass type Hamiltonian
\begin{equation}
\Hlogicf = \sum_{i=1}^{N-1} \sum_{j=i+1}^{N} J_{i,j} \, \paulilogic_z^{(i)} \paulilogic_z^{(j)}
\label{eq:all_to_all_hamiltonian}
\end{equation}
where the interactions $J_{i,j}$ encode some problem we wish to solve by finding the lowest energy configuration, and $i$, $k$ run over spin-1/2 sites.
For the purposes of this paper, we choose $J_{i,j}$ from the binary configurations $\{0,1\}$, which are already sufficient to encode a range of hard problems, such as the well known \textsc{max-cut} \cite{maxcut} problem. Here, we follow LHZ by choosing $J_{i,j} \sim U(-J, J)$, i.\ e.\ the interaction matrix elements being continuously, uniformly sampled with $J=1$.

We can obtain a quantum annealer by adding, for instance, a local transverse field
\begin{equation}
\Hlogici = \sum_i \paulilogic_x^{(i)}
\end{equation}
for which the initial ground state can be easily prepared,
and then introduce two ramps $A(\tau)$ and $B(\tau)$ as follows
\begin{equation}
\Hlogic{(\tau)} = \Alogic{(\tau)} \Hlogici + \Blogic{(\tau)}\Hlogicf.
\end{equation}
In the adiabatic (coherent) quantum annealing approach, the initial part is slowly ramped down from $\Alogic{(\tau=0)}=1$ to $\Alogic{(\tau=1)}=0$, while the final part is ramped up, $\Blogic{(\tau=0)}=0, \Blogic{(\tau=1)}=1$, over a timescale $T$ which is large enough to allow an adiabatic passage such that at $t=T$ ($\tau=1$) we arrive in the ground state of $\Hlogicf$.

\paragraph{LHZ architecture:}

We obtain the LHZ architecture by mapping the parity of every pair of two \textit{logical} spins $(i,j)$ from the fully connected, original problem, onto a single \textit{physical} spin $k$. 

As a result, from $N$ \textit{logical} spins, we obtain $K=N(N-1)/2$ \textit{physical} spins with local fields $J_k = J_{i,j}$, and can directly map the final Hamiltonian onto
\begin{equation}
\Hphysf = \sum_{k=1}^K J_{k} \, \pauliphys_z^{(k)}
\label{eq:problem_hamiltonian}
\end{equation}
where $\pauliphys_x$, $\pauliphys_z$ are the spin-1/2 Pauli operators acting on the \textit{physical} spins.
The mapping is shown in Fig.~\ref{fig:architecture}. 
However, it is only invertible upon introduction of at least $K-N$ constraints mapping from $K$ \textit{physical} qubits back onto the $N$ \textit{logical} qubits.
In the LHZ architecture, the \textit{logical} state is resolved only up to global spin-flip parity, by introduction of $N_c = K-N+1$ constraints: The number of spin up must be even in every plaquette $p$. These constraints can be realized in the ground state by introducing an energy penalty of the form
\begin{equation}
\Hphyscplaq^{(p)} = - \pauliphys_z^{(p_1)}\pauliphys_z^{(p_2)}\pauliphys_z^{(p_3)}\pauliphys_z^{(p_4)}
\label{eq:energy_penalty}
\end{equation}
where $p_1\dots p_4$ label the spin in the four corners of plaquette $p$. 
Again, for an annealing process, we introduce a simple local initial Hamiltonian
\begin{equation}
\Hphysi = \sum_{k=1}^K \pauliphys_x^{(k)}.
\label{eq:initial_hamiltonian}
\end{equation}
The full passage Hamiltonian then reads
\begin{equation}
\Hphys{(\tau)} = \Aphys{(\tau)} \Hphysi + \Bphys{(\tau)} \Big (\Hphysf + \sum_{p=1}^{N_c} C^{(p)}{(\tau)} \Hphyscplaq^{(p)} \Big).
\label{eq:passage_hamiltonian}
\end{equation}
Aside from the ramps $\Aphys{(\tau)}, \Bphys{(\tau)}$, we have introduced time-dependent constraint strengths $C^{(p)}$ which can be tuned freely during the adiabatic passage but must ultimately become large enough to make sure the ground state fulfills all constraints at $\tau=1$. This preserves the one-to-one mapping between \textit{logical} and \textit{physical} qubits. As LHZ point out, a larger (possibly uniform) constraint strength $C^{(p)}$ also increases the value of the minimal gap in the \textit{physical} regime, and thereby adds to the discrepancy of minimum gaps between both regimes. There is however a sweet spot in the functions $C^{(p)}$, which are sufficiently large for the constraints to stay intact at the end of the sweep, and also suppress a significant increase in the minimum gap within the \textit{physical} LHZ scheme. In the supplementary material of the work by Lechner et al., the authors give quantitative suggestions for $C^{(p)}$ for $N \in \{3,4\}$ \textit{logical} qubits.

\section{Design of fixed optimized schedules}
\label{sec:optimized_schedules}

The rate at which a system can be evolved in time is critically dependent on the magnitude of the minimum gap (and its occurence within the sweep). As we vary the interaction matrix in Eq.~\ref{eq:problem_hamiltonian}, we consequently change its energy landscape. Two Ising problems with arbitrary interaction matrices may exhibit, for instance, similar energy landscapes and thus energy gaps. Hence, the speed at which adiabatic time evolution can be performed should exhibit a comparable upper bound.  Systems of ten or fewer \textit{physical} qubits do hardly exhibit multiple minima in our training sample: For $K=10$ we find the ratio of instances that have at least one local minimum alongside a global minimum is $<1$\%. 
Moreover, as can be seen from Fig.~\ref{fig:N_5_gaps_and_ramps}, the time of the emergence of the avoided crossing within the sweep is correlated to the magnitude of $\Delta E$. The larger $\Delta E$ becomes, the earlier the emergence of the avoided crossing tends to occur. Since time of occurrence of the minimal gap is linked to its magnitude, and the number of instances exhibiting local minima is insignificant, we focus on the amplitude of the minimal gap as our main metric used to generate problem groups.

The groups are designed such that if two instances are taken from the same group, they will exhibit a comparable minimum gap value. If one instance is taken from one group $g_i$, and another instance is taken from another group $g_{i'}$, $i \neq i'$, then the discrepancy between the two energy gaps grows. Consequently, for any two instances taken from $g_i$, we expect the annealing time required for a successful transition into the desired output state to be of similar order. This motivates the search for finite set of annealing protocols, where the idea is to customize protocols according to the different groups.

\subsection{Problem groups}
We start from a large sample $M$ containing $\lvert M \lvert = 4\cdot10^4$ programmable instances, which are used to construct a training set of in total $2400$ instances. Implementation and sampling details can be found in appendix~\ref{sub:implementation_details}. The first step is to perform an exact diagonalization on all instances in order to obtain their individual minimum gap $\Delta E$. We then sort all instances according to $\Delta E$, such that for all instances $j \in M$ 
\begin{equation}
\Delta E_j \leq \Delta E_{j+1}
\label{eq:sorted_gaps}
\end{equation}
applies. Hence, the first instance ($j=1$) has the smallest minimum gap, whereas the instance indexed by $j=\lvert M \lvert$ has the largest gap. The next step is to choose $\lvert G \lvert$, i. e. the number of groups included in $G$. In this paper, the authors chose the number of groups to be six. This was motivated by the notion, that less groups make the question of how large to choose $T$ for all instances more difficult to solve, while too many groups render the applicability of all protocols in $S$ too costly. 

\begin{figure}
\centering
\includegraphics[scale = 0.17]{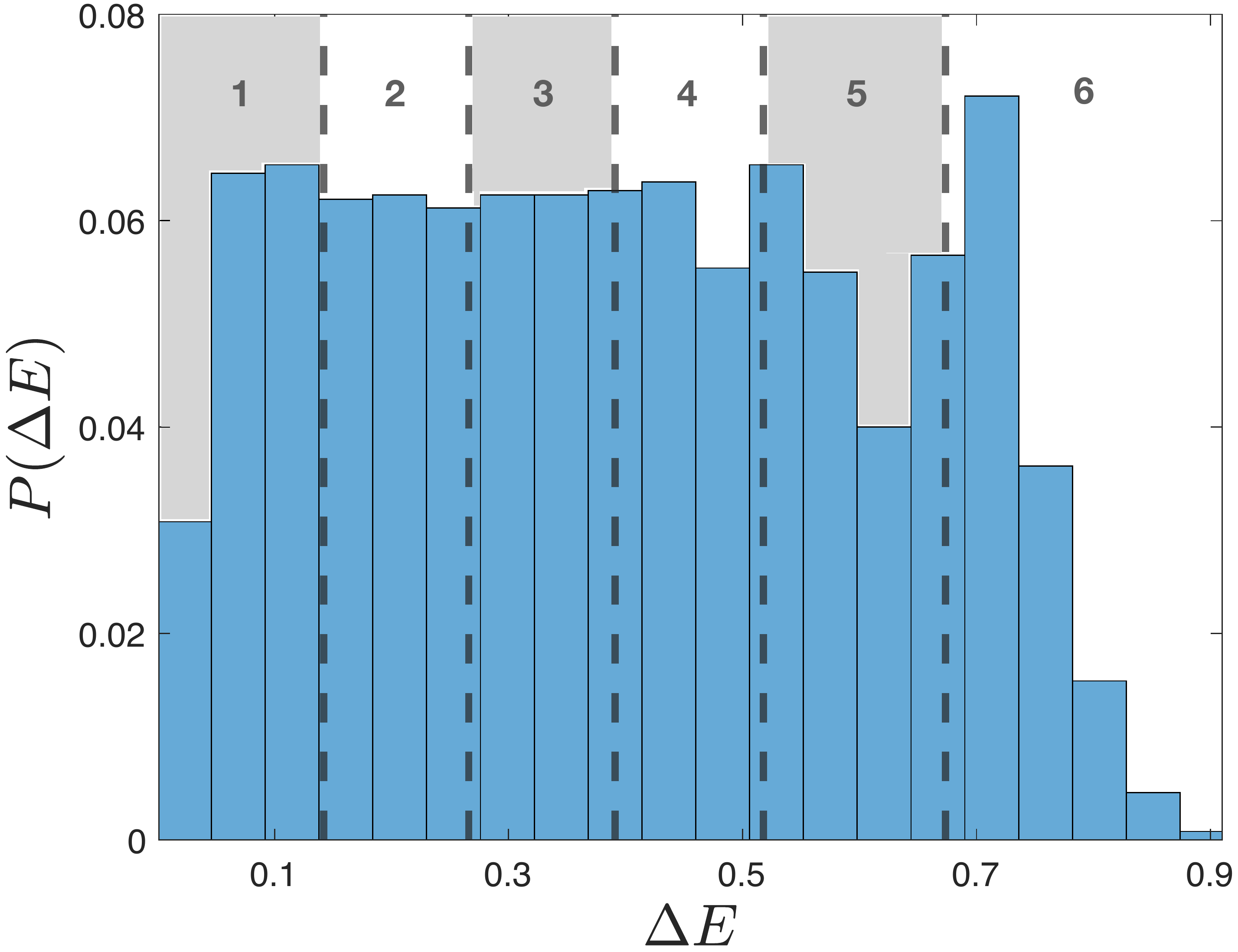}
\caption{(Color online) Histogram of minimum gaps. Areas of alternating shades represent all instances within a group $g_i$, numbered in ascending order at the top. Their width indicate the range of values in $\Delta E$ for each group.}
\label{fig:N_5_categories_dE}
\end{figure}

Afterwards, we put the first $\lvert S\lvert/ \lvert G \lvert$ instances in $g_1$ and note the standard deviation $\sigma_1(\Delta E_1, \cdots, \Delta E_{\lvert S\lvert/\lvert G \lvert})$ in terms of their energy gaps. We do this for all groups in $G$. In this manner, we are able to assign a standard deviation $\sigma_i$ to each group $g_i$. The goal is to shift instances from $g_i$ to neighbored groups $g_{i\pm 1}$ such that (i) $\sigma_1 \approx \sigma_2 \approx \cdots \approx \sigma_{\lvert G \lvert}$ and (ii) $\sigma_i$ is as small as possible for all $i$. This approach has one major advantage: all $\Delta E$ from individual instances in a given group $g_i$ are as similar as possible. This implies, that each group $g_i$ is maximally different from the other groups $g_{i' \neq i}$. Hence, applying a group specific optimized protocol $s_i(\tau)$ onto any instance $j$ in $g_i$ yields meaningful results as opposed to applying it to an instance $j'$ in $g_{i'}$, $i \neq i'$ and $j \neq j'$. At this point six groups spanning unique intervals of minimum energies exist. Our analysis suggests that 400 instances per group are sufficient to represent it fully (i. e. 2400 instances in total), which is on par with the analysis carried out in \cite{Lechner2015annealer}, where in total 400 instances were analyzed. Trimming down the number of instances per group to $\lvert g_i \lvert = 400$ leaves us with the final groups accompanied by the distribution of $\Delta E$, Fig. \ref{fig:N_5_categories_dE}.

The histogram of instances in the training set $G$ is illustrated in Fig. \ref{fig:N_5_categories_dE}. Groups of instances $g_i$ are shaded in grey and white. Their width span the interval of $\Delta E$ for all instances within their corresponding group.

\subsection{Optimized schedules}

Given a set $G$, we are now able to find out the set of corresponding optimized annealing protocols $S$ by employing the dCRAB algorithm. As mentioned in Section \ref{sec:intro}, the first step is to define $A(\tau) \equiv 1-s(\tau)$ and $B(\tau) \equiv  s(\tau)$, Eq. \ref{eq:passage_hamiltonian}. Due to the simple relation between $A(\tau)$ and $B(\tau)$, it is sufficient to denote the set of protocols $S=\{s_1(\tau), s_2(\tau), dots\}$ as mentioned in Section \ref{sec:intro}. Moreover, we omit the explicit time-dependence of the protocols $s_i(\tau)$ to shorten equations and refining the figures, except when it serves the purpose of clarity. From here on, optimized protocols are called $s_i$, where $i$ indicates the corresponding group, the protocol was designed for. Furthermore, due to the simple relation $A(\tau) = 1-B(\tau)$, it is sufficient to denote the set of protocols as $S=\{s_1, s_2, \dots\}$. To measure the effect of a protocol $s_i$ onto the time-evolution of all instances (indexed as $j$) in $g_i$, we take the average group fidelity $\mathcal{F}(s_i,g_i)$ into account, denoted by
\begin{equation}
\mathcal{F}(s_i,g_i) = \frac{1}{\lvert{g_i}\lvert}\sum_{j \in g_i}{\mathcal{F}(s_i,j)}.
\label{eq:fidelity}
\end{equation}
It is the average ground state fidelity  at $\tau=1$, which will be the figure of merit in the optimization loop. Here, the fidelity for a single instance $j$
\begin{equation}
\mathcal{F}(s_i, j) = \lvert \langle \Psi_\text{sim}(s_i, j) \lvert \Psi_\text{exact}(j) \rangle \lvert^2
\label{eq:single_instance_fiddy}
\end{equation}
is the overlap of simulated state $\lvert \Psi_\text{sim}(s_i, j) \rangle$ and exact ground state $\lvert \Psi_\text{exact}(j) \rangle$ of the final Hamiltonian $\Hphysf$, Eq.~\ref{eq:problem_hamiltonian}, respectively.

\begin{figure}
\centering
\includegraphics[scale = 0.18]{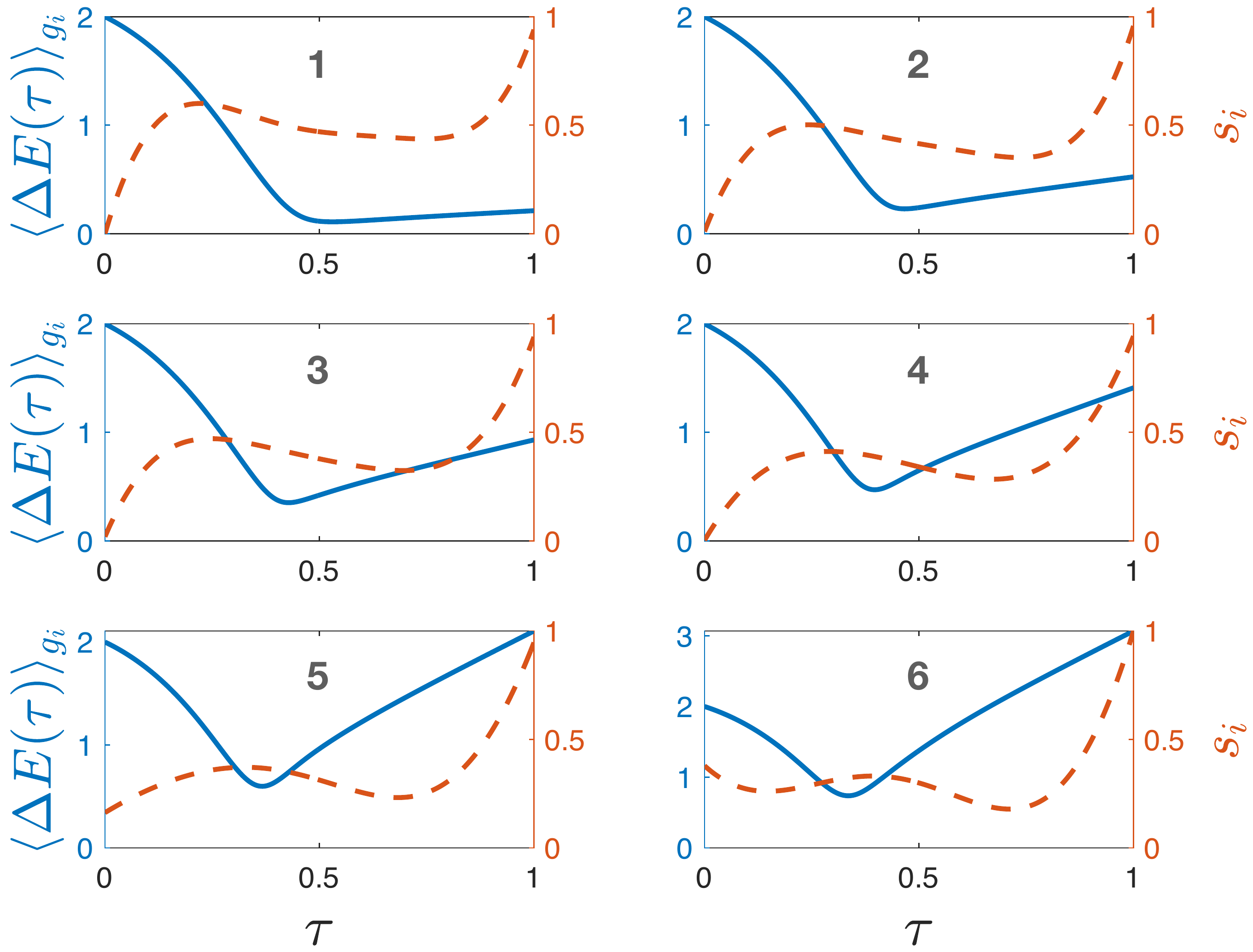}
\caption{(Color online) Instantaneous average energies $\langle \Delta E(\tau)\rangle_{g_i}$ averaged over all instances in $g_i$ (solid line) along with their corresponding optimized annealing protocols in $s_i$ (dashed line) numbered in ascending order at the top.}
\label{fig:N_5_gaps_and_ramps}
\end{figure}

The goal fidelity is chosen to be $\mathcal{F}(s,g_i) \geq 0.9$. The choice of this specific number is motivated in Appendix \ref{sec:justification_goal_fidelity}.
As the group-specific average minimum gap $\Delta E_{g_i}$ heavily influences the required simulation time $T$ to reach a given target ground state fidelity $\mathcal{F}(s_i, g_i) > 0.9$, we increase $T$ until the target fidelity is reached. Fig.~\ref{fig:N_5_gaps_and_ramps} shows for all groups $g_i, i \in \{1, 2, 3, 4, 5, 6\}$ the average energy gap $\Delta E_{g_i}$ for all instances in $g_i$ along with the corresponding optimized schedule $s_i$.

We report that when the average minimum gap $\Delta E_{g_i}$ over all instances in $g_i$ becomes smaller, then the time of its occurrence (sometimes referred to as position) shifts to the first half of the sweep. The protocols $s_i$ are reminiscent of characteristics from an optimal adiabatic ramp \cite{Roland_2002} as the first time derivative of $s_i(\tau)$ reaches a saddle-point, when the energy gap gets close to its minimum.

\section{Performance analysis}
\label{sec:results}

\begin{figure}
\centering
\includegraphics[scale = 0.19]{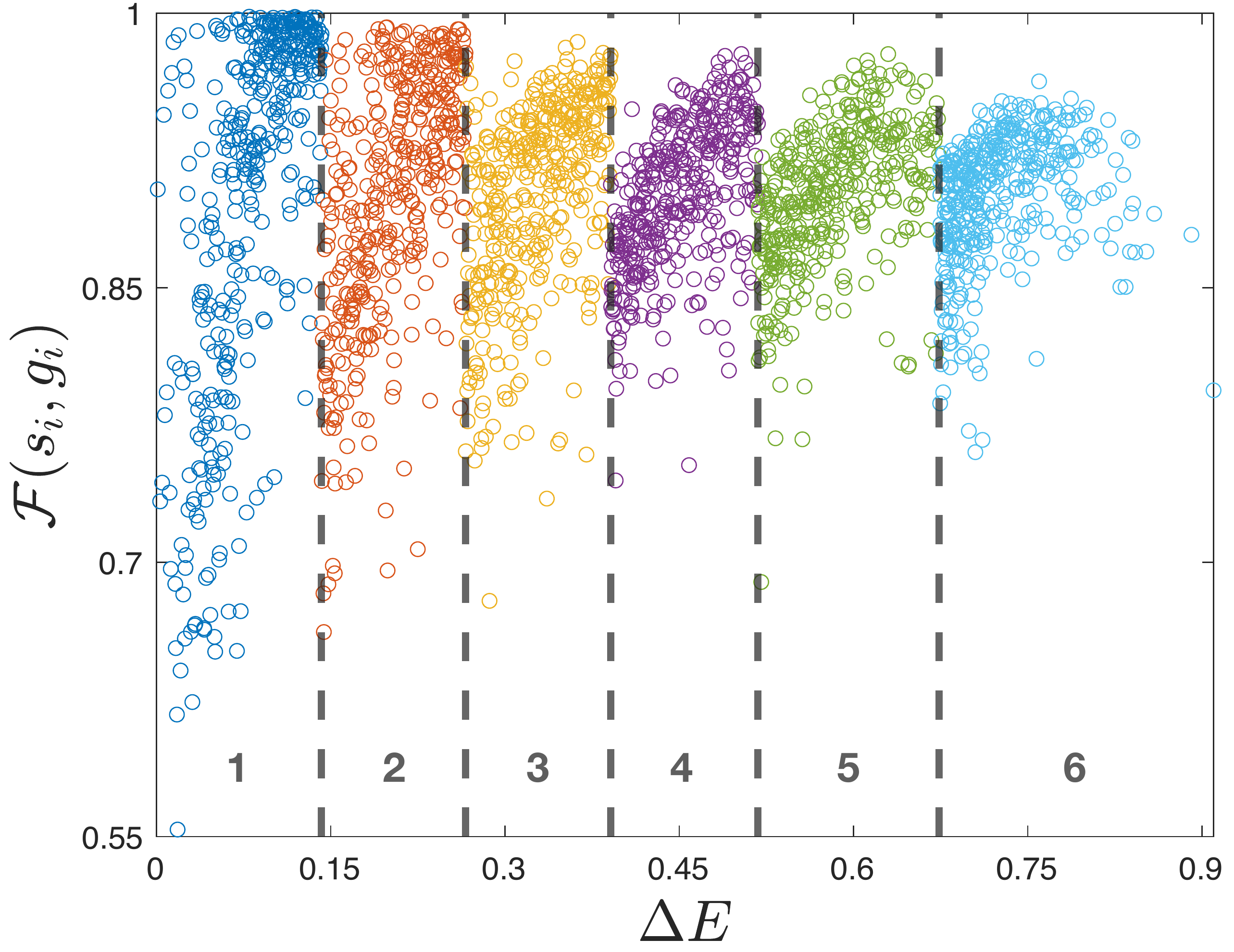}
\caption{(Color online) Single-instance fidelities $\mathcal{F}(s_i, j \in g_i)$ from optimal protocols $\{s_i\}$ onto all instances $j$ of corresponding groups $\{g_i\}$ separated by solid vertical lines and numbered in ascending order at the bottom.} 
\label{fig:opt_schedule_N_5}
\end{figure}

\begin{figure}
\centering
\includegraphics[scale = 0.45]{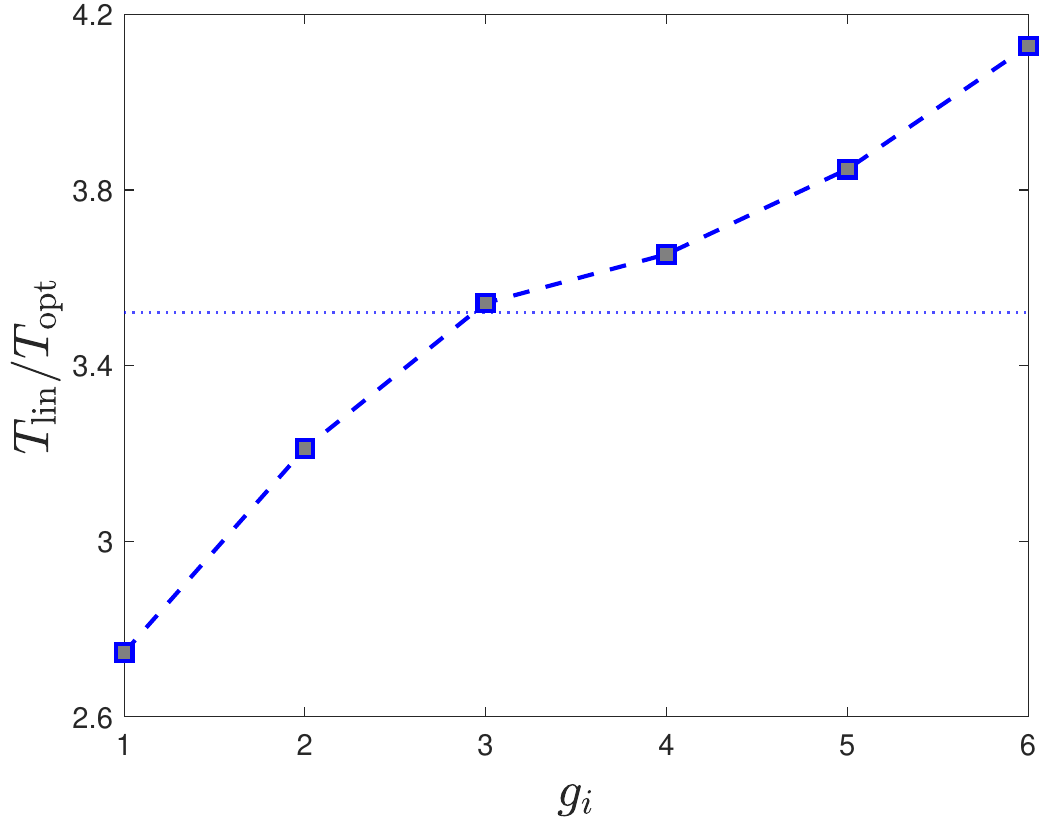}
\caption{(Color online) Simulation time speed-up for $N=5$ from the set of optimized protocols $s_i$ relative to a linear ramp to reach mean goal fidelity $\mathcal{F}(s_i,g_i) \geq 0.9$. Thick lines are group specific speed-ups, thin lines represent average speed-up of factor $\sim 3.52$.}
\label{fig:speed_up}
\end{figure}

We show the distribution of single-instance fidelities $\mathcal{F}(s_i,j)$, Fig. \ref{fig:opt_schedule_N_5}, when the optimized protocol $s_i$ is applied to its corresponding instances $j$ in $g_i$. Vertical lines separate the groups in ascendending order from left to right. Clearly, the smaller the gaps become, the wider the spread of single-instance fidelities. This is due to the fact that the smaller the average energy gap per group, the wider is the range of reciprocal energies, which impacts the speed limit and fidelity directly. Another reason is that in $g_1$, minimum gaps can be arbitrarily close to zero, leading to an explosion in required simulation time $T$ as per QSL, which in turn might be an indication of hardness. We find the ratio of instances with more than one energy minimum to be $<1\%$ for $N=5$ \textit{logical} qubits. It shall be pointed out, that for system sizes of $N>5$, multiple local are more likely to occur \cite{Tanaka1980}, contributing further to the hardness of the problem.
Next, we compare required annealing times of linear and optimized annealing protocols, Fig. \ref{fig:speed_up}, for reaching the mean goal fidelity $\mathcal{F}(s_i,g_i) \geq 0.9$. Typically, speed-up becomes more meaningful for larger systems as $\Delta E$ rapidly shrinks for larger systems,~\cite{Albash2018}. For $N=5$ we have an averaged reduction of annealing time of $71.6\%$, i. e. one needs 3.52 fold larger annealing times with linear protocols compared to optimized ones.

We also probe optimized protocols in $S$ gathered by analysis of the training set onto a different test set $G^\text{test}$ of again $6 \cdot 400$ instances. Albeit no optimizations were performed with respect to the test set, a tiling into different groups $g_i^\text{test}$ is required to verify, that the protocols are applicable and useful for any transverse-field Ising instance within the LHZ model and not a product of overfitting. 

\begin{figure}
\centering
\includegraphics[scale = 0.47]{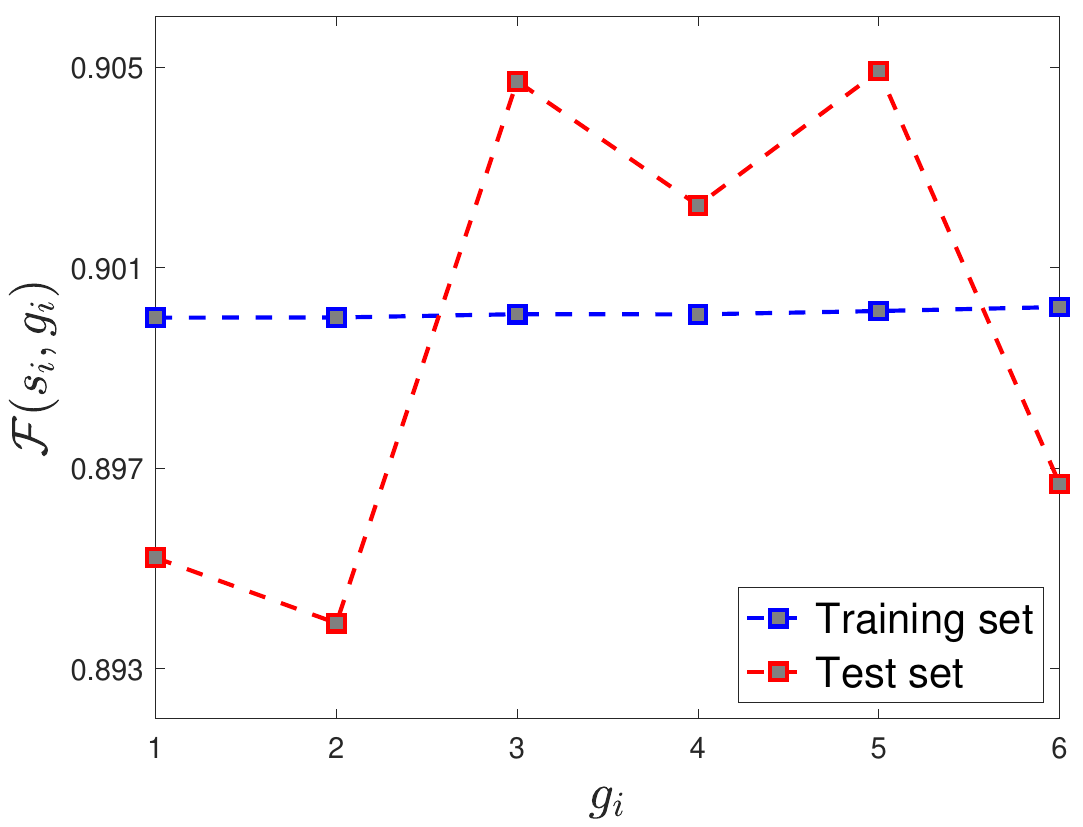}
\caption{(Color online) Comparison of averaged fidelities $\mathcal{F}(s_i,g_i)$ and $\mathcal{F}(s_i,g^\text{test}_i)$ for $N=5$ when applying optimized protocols in \{$s_i$\} to all instances $j$ in the training set \{$g_i$\} and test set \{$g^\text{test}_i$\}. By design, $\mathcal{F}(s_i,g_i)$ is just above the 0.9 line.}.
\label{fig:N_5_fidelity_comparison}
\end{figure}
\begin{figure}
\includegraphics[scale = 0.19]{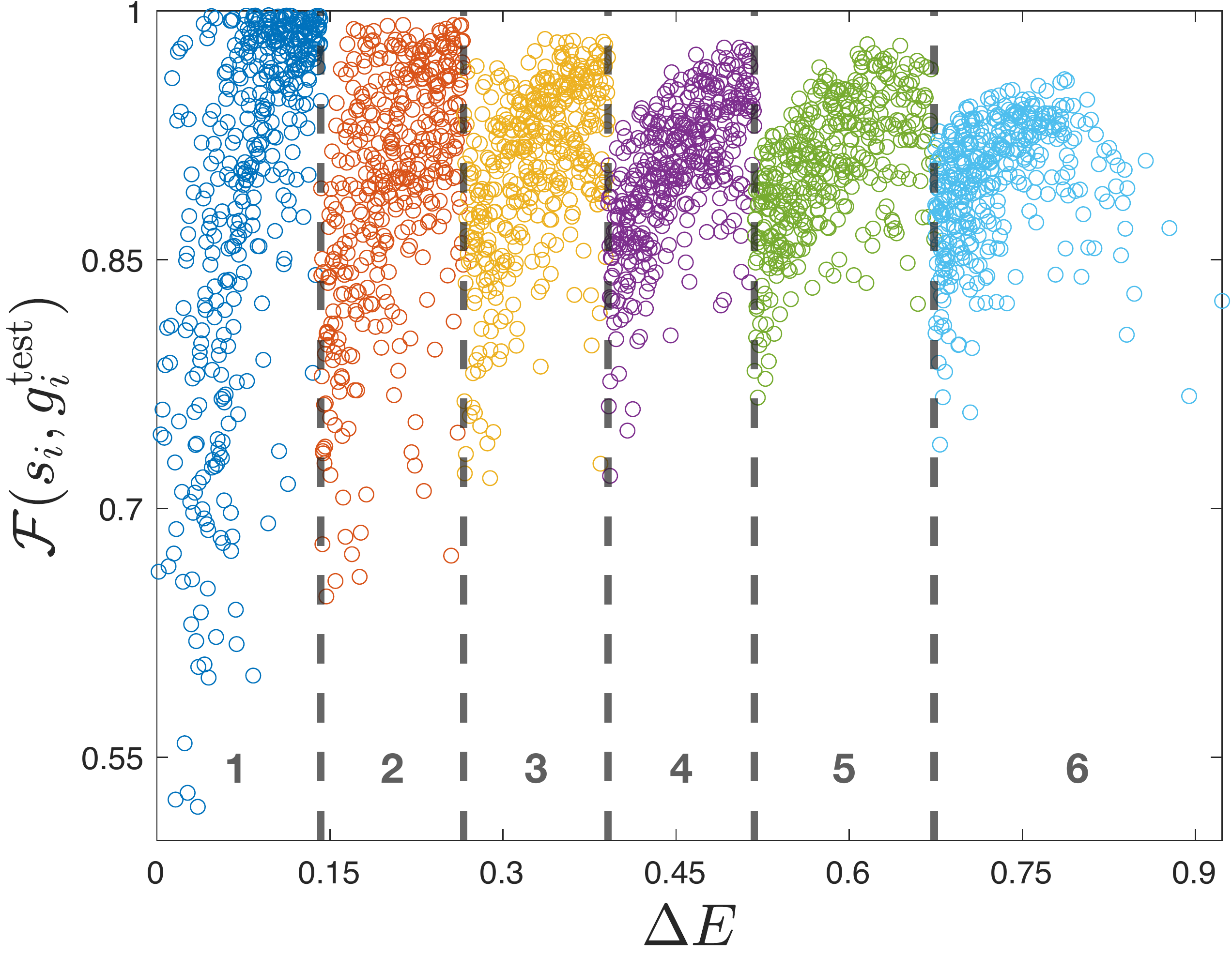}
\caption{(Color online) Single-instance fidelities $\mathcal{F}(s_i, j \in g^\text{test}_i)$ in the test group from optimal protocols $\{s_i\}$ onto all instances $j$ of corresponding groups $\{g^\text{test}_i\}$ separated by solid vertical lines and numbered in ascending order at the bottom.}.
\label{fig:N_5_fidelity_comparison_test_group}
\end{figure}

We find that $\forall i \in \{1,2,3,4,5,6\}$ it holds $\mathcal{F}(s_i,g_i)\approx\mathcal{F}(s_i,g^\text{test}_i)$. This indicates that optimized protocols $s_i$ indeed cover a wide range of programmable LHZ problems and can successfully be applied to arbitrary instances sampled via a similar distribution in $J$, cf. Appendix \ref{sub:implementation_details}. Zooming in, cf. Fig. \ref{fig:N_5_fidelity_comparison_test_group}, we find that the lowest data point in fidelity $\mathcal{F}(s_i, g^\text{test}_i)$ is still above the 0.5 line.

Investigations for larger systems ($N>5$) need to be carried out in order to probe the usability of the method and whether or not optimized protocols can be found, not to mention the number of groups necessary to cover a wide range of programmable LHZ problems. After all, larger systems are typically more relevant for real world applications, but often come with multiple local energy minima \cite{schnabel}, which makes the design process of groups more complicated. Larger systems may pick up a Stückelberg phase when passing through multiple avoided crossings~\cite{Shevchenko_2010}, which disturbs the single-instance wave function $\lvert \Psi_\text{sim}(s, j) \rangle$ and thereby fidelity, Eq.~\ref{eq:single_instance_fiddy}, in a non-trivial way. 

\section{Conclusion}
\label{sec:conclusion}
On the LHZ annealing architecture ~\cite{Lechner2015annealer}, our stated goal is to provide a set $S$ consisting of fixed optimized annealing protocols $s(\tau)$, that outperforms linear passage on a large class of transverse-field Ising problems. In doing so, we analyzed a large set of generated instances in terms of their system dynamics.
A minimum time is required in order to convert two quantum states. For the considered system size $N=5$ this required time is critically dependent on the minimum energy gap $\Delta E$ between ground state and first excited state within the instantaneous energy spectrum. Therefore, six groups $g_i, i = 1,...,6,$ were established, that collect instances with similar traits, specifically, with comparable $\Delta E$ per instance. This way, we can adjust required annealing times for each individual group within $g_i$, paving the way for an optimization over all instances $j$ $\in g_i$ $\forall i \in \{1,2,3,4,5,6\}$. What we find are six optimized protocols $s_i(\tau), i = 1,...,6$, that if applied to its corresponding group $g_i$ yields an average ground state fidelity of $\geq 0.9$. Using these protocols for their corresponding groups $g_i$ reduces required simulation time by an average of 71.6\% relative to linear protocols.
Although the method is not specific to the LHZ model and could in principle be applied to larger systems with local couplings, we chose to work within the LHZ framework, mapping $N$ \textit{logical} qubits onto $K=N\cdot (N-1)/2 = 10$ \textit{physical} qubits. Thereby, we provide a framework to generate intermediate optimized protocols ready to be used by NISQ annealers, that utilize local qubit couplings. One Ansatz to overcome the drawback of larger system sizes in the $physical$ regime and yet keep the benefit of applicability would be to find a functional $F$ that takes as input the optimized protocol for the \textit{logical} problem ($\mathcal{H}=2^N$) and puts out the optimized schedule for the \textit{physical} problem ($\mathcal{H}=2^K$), i. e. $F[s_i^\text{logical}(\tau)] \mapsto s_i^\text{physical}(\tau)$.

Note, that larger systems exhibit multiple local minima. In this case, we should not only group instances according to their $\Delta E$, but also according to their number of local minima. Further investigations are needed as to how many groups are required for a beneficial tradeoff between correctness and cost. More groups mean more trial protocols. This decreases applicability, but naturally increases the performance of protocols.

As a last remark we point out, that optimized protocols as presented in this paper require knowledge of the minimum gaps, which were found via exact diagonalization. As this strategy is not applicable in a laboratory, we explore a method to generate optimized protocols on real annealers in Appendix \ref{sec:laboratory}. Future work is comprised of executing this strategy and explore the adiabaticity of dynamics. Single-instance protocol optimizations will violate the adiabatic theorem for too short annealing times. Thereby, the system may be forced into a diabatic evolution, successively jumping between energy levels. In contrast to adiabatic evolution, at which the ground state population stays constant, diabatic evolution implies that population transfer follows a more complex pattern. This can in principle be exploited, given the number of protocols in $S$ is much smaller than the number of problems it covers. As optimized protocols within an experimental setup are based on single-instance protocol optimizations, each protocol has a dedicated parent problem which serves as an archetype representing a specific energy landscape. On the one hand, there is still the need to use techniques to get the energy landscape of the archetypical instance. On the other hand, if we had full spectral knowledge on the problems which lay the foundation to the set of optimized protocols, we may be able to deduce from that the energy landscapes of problems, which respond favorably to a particular protocol. Thus, once we are capable of efficiently extracting information on instantaneous energy levels of various different problems, the sole application of protocols can potentially proxy the identification of energy landscapes of an underlying problem instance.

\section*{Acknowledgements}
The authors gratefully acknowledge support and computational ressources from Mercedes-Benz AG, OpenSuperQ (820363) and German Federal Ministry of Education and Research
in the funding program “quantum technologies – from
basic research to market” (contract number 13N15582). Furthermore, we acknowledge useful conversations with Ferdinand Tschirsich, Timo Felser, Marco Rossignolo, Thomas Reisser, Tim Bode, Dmitry Bagrets and David Headley.

\newpage

\bibliography{manuscript}

\appendix
\section{Implementation details}
\label{sub:implementation_details}
We generated transverse-field Ising models according to $\Hphysf$, cf. Eq. ~\ref{eq:problem_hamiltonian}. Specifically, we varied the interaction matrix $J$ in the uniform interval [-1,1]. Randomization was performed by a Mersenne-Twister python implementation. The constraints $C$ are equal for all plaquettes $p$ and ramped up linearly as $C(\tau)=\tau \cdot C$, i. e. decoupled from the problem Hamiltonian (Eq. \ref{eq:passage_hamiltonian}). The strength was chosen according to analysis by LHZ for $N \in \{3,4\}$. For five \textit{logical} qubits we set $C \equiv C^{(p)} = 2.0$. 

Transverse-field Ising problems, which exhibit a ground state degeneracy at or critically close to $t \approx T$ were discarded for simplification of ground state fidelity evaluations. If the degeneracy appears at $t=T-\delta$, and $\delta$ is in the scale of the smallest numerical time step, then the system has hardly time to emancipate from its degeneracy. Additionally, some instances did violate the constraints and thus prevent a one-to-one mapping between \textit{logical} and \textit{physical} qubits. Lastly, instances for which a fidelity $\mathcal{F} \geq 0.9$ can not be reached within a reasonably large simulation time $T=1000$ were discarded.

The training set consists of $6\cdot 400$ instances, whereas we start from in total $4 \cdot 10^4$ programmable problems. Constructing six groups leaves us with uneven numbers of instances per group. How many instances do we need to represent a specific group? LHZ used in total 400, and we find that a histogram over the distribution of $\Delta E$ is not harmed by carefully cutting off instances, such that we deem 400 instances per group sufficient to solve the stated goals.

\section{Justification of goal fidelity}
\label{sec:justification_goal_fidelity}
We justify the goal fidelity of $\mathcal{F}(s_i, g) > 0.9$ via the following arguments.
\begin{enumerate}
  \item The required fidelity drastically depends on the annealing time $T$. Hence, as we fixate the goal fidelity to 0.9, we can perform the task of finding optimized schedules much quicker than if we were searching for fidelities close to unity.
  \item Some instances in $g_1$ demand excessive computational resources to be treated in an adiabatic way. For those instances, we find the minimum gap $\Delta E$ to be close to zero and the emergence of the avoided crossing to be around $\tau \lesssim 1$. This might indicate an artifact of too small choice of $C^{(p)}$. Still, as such instances contribute to the average fidelity, this limits the upper bound of fidelity values that can be reached.
  \item The protocols in $\{S\}$ are by design not optimal for any arbitrary single instance, but can be used as guess pulses for single-instance protocol optimizations. If we were to target group-specific average fidelities of $\mathcal{F}(s,g_i)>0.99$, we might face the risk of over-optimizing protocols and thereby hamper the single-protocol optimization for arbitrary LHZ instances.

\end{enumerate}

\section{Optimized protocols from annealing experiments}
\label{sec:laboratory}
Minimum gaps are not readily available in a real hardware environment, which means that the strategy proposed in this paper can not be applied directly. However, as proof of concept the work shows that in principle finding optimized protocols is possible that work well for a large class of programmable LHZ problems. 
The following method not only can be readily applied on annealers. It also alleviates the problem of guessing the optimal number of groups as well as the optimal number of instances within these groups as these quantities are output as byproducts when executing the method.

The first step is to define two threshold probabilities that proxy ground state fidelities, e. g. $\mathcal{F}_- \equiv 0.66$ and $\mathcal{F}_+ \equiv 0.9$. Next, we sample a sufficiently large number of LHZ instances (or Ising spin glass instances) and draw from this sample the first instance and perform an energy optimization on an annealer. Again, the annealing time needs to be varied until with large certainty $\geq \mathcal{F}_+$ the minimum energy state of a final Hamiltonian is reached. Step two is to draw the second instance and apply the previously found protocol with corresponding annealing time. If we do not find with lower bound certainty $\geq \mathcal{F}_-$ the minimum energy state, a single-protocol optimization is again performed until with probability $\geq \mathcal{F}_+$ we find the minimum energy state. While in the first case we simply go to the next instance and apply the one found protocol, in the second case we have now two schedules to be applied to the successor. This scheme is repeated until the number of protocols with corresponding annealing times saturates. As energy landscapes become increasingly complex with larger systems, diabatic evolution may become unique for each individual instance. The sample size and number of protocols are then difficult to anticipate --- in the worst case the number of protocols scales linearly in the number of instances.

\end{document}